\input harvmac
\input epsf
\noblackbox
\newcount\figno
\figno=0
\def\fig#1#2#3{
\par\begingroup\parindent=0pt\leftskip=1cm\rightskip=1cm\parindent=0pt
\baselineskip=11pt
\global\advance\figno by 1
\midinsert
\epsfxsize=#3
\centerline{\epsfbox{#2}}
\vskip 12pt
\centerline{{\bf Figure \the\figno} #1}\par
\endinsert\endgroup\par}
\def\figlabel#1{\xdef#1{\the\figno}}
\def\pano{\par\noindent}

\font\cmss=cmss10
\font\cmsss=cmss10 at 7pt
\def\half{{1\over 2}}
\def\rlx{\relax\leavevmode}
\def\inbar{\vrule height1.5ex width.4pt depth0pt}
\def\IC{\relax\,\hbox{$\inbar\kern-.3em{\rm C}$}}
\def\IR{\relax{\rm I\kern-.18em R}}
\def\IN{\relax{\rm I\kern-.18em N}}
\def\IP{\relax{\rm I\kern-.18em P}}
\def\ZZ{\rlx\leavevmode\ifmmode\mathchoice{\hbox{\cmss Z\kern-.4em Z}}
 {\hbox{\cmss Z\kern-.4em Z}}{\lower.9pt\hbox{\cmsss Z\kern-.36em Z}}
 {\lower1.2pt\hbox{\cmsss Z\kern-.36em Z}}\else{\cmss Z\kern-.4em Z}\fi}
\def\narrowplus{\kern -.04truein + \kern -.03truein}
\def\narrowminus{- \kern -.04truein}
\def\narrowminussub{\kern -.02truein - \kern -.01truein}

\def\a{\alpha}
\def\ts{\textstyle}

\def\o#1{\overline{#1}}


\lref\BLDI{J. Blum and K. Dienes,  {\it Strong/Weak Coupling Duality Relations 
for Non-supersymmetric String Theories}, Nucl.Phys. {\bf B516} (1998) 83,
hep-th/9707160.} 

\lref\RTEV{I. Antoniadis, {\it A Possible New Dimension at a Few TeV},
Phys.Lett. {\bf B246} (1990) 377; \hfil\break
J. Lykken, {\it Weak Scale Superstrings}, 
Phys.Rev. {\bf D54} (1996) 3693, hep-th/9603133; \hfil\break
N. Arkani-Hamed, S. Dimopoulos and G. Dvali, 
{\it The Hierarchy Problem and New Dimensions at a Millimeter}, 
Phys.Lett. {\bf B429} (1998) 263, hep-ph/9803315; \hfil\break
N. Arkani-Hamed, S. Dimopoulos and G. Dvali,
{\it Phenomenology, Astrophysics and Cosmology of Theories with Sub-Millimeter
       Dimensions and TeV Scale Quantum Gravity},
Phys.Rev. {\bf D59} (1999) 086004, hep-ph/9807344; \hfil\break
I. Antoniadis, N. Arkani-Hamed, S. Dimopoulos, G. Dvali, {\it
New Dimensions at a Millimeter to a Fermi and Superstrings at a TeV},
Phys.Lett. {\bf B436} (1998) 257, hep-ph/9804398; \hfil\break
 G. Shiu, S.-H. H. Tye, {\it TeV Scale Superstring and Extra Dimensions},
Phys.Rev. {\bf D58} (1998) 106007, hep-th/9805157.}

\lref\BB{I. Antoniadis, E. Dudas and A. Sagnotti, {\it Brane
Supersymmetry Breaking}, Phys. Lett. {\bf B464} (1999) 38, 
hep-th/9908023; \hfil\break
C. Angelantonj, I. Antoniadis, G. D'Appollonio, E. Dudas,
A. Sagnotti, {\it Type I Vacua with Brane Supersymmetry Breaking},
hep-th/9911081.}

\lref\URA{G. Aldazabal and A.M. Uranga, {\it Tachyon-free
Non-supersymmetric Type IIB Orientifolds via Brane-Antibrane Systems}, 
JHEP {\bf 9910} (1999) 024, hep-th/9908072; \hfil\break
G. Aldazabal, L.E. Ib\'a\~nez and  F. Quevedo, {\it Standard-like
Models with Broken Supersymmetry from Type I String Vacua}, JHEP 
{\bf 0001} (2000) 031, hep-th/9909172.}

\lref\BGKL{R. Blumenhagen, L. G\"orlich, B. K\"ors and D. L\"ust,
    {\it  Noncommutative Compactifications of Type I Strings on Tori with 
    Magnetic Background Flux},  JHEP {\bf 0010} (2000) 006,  hep-th/0007024;
    \hfil\break
    R. Blumenhagen, L. G\"orlich, B. K\"ors and D. L\"ust,  
   {\it Magnetic Flux in Toroidal Type I Compactification},
    hep-th/0010198.}

\lref\ABG{C. Angelantonj, R. Blumenhagen and M.R. Gaberdiel,
{\it Asymmetric Orientifolds, Brane Supersymmetry Breaking and Non BPS Branes},
Nucl.Phys. {\bf B589} (2000) 545, hep-th/0006033;  \hfil\break
R. Rabadan, A.M. Uranga, 
 {\it Type IIB Orientifolds without Untwisted
  Tadpoles, and non-BPS D-branes}, hep-th/0009135.}

\lref\AADS{C. Angelantonj, I. Antoniadis, E. Dudas and A. Sagnotti, {\it
       Type-I Strings on Magnetised Orbifolds and Brane Transmutation},
       Phys.Lett. {\bf B489} (2000) 223, hep-th/0007090;  \hfil\break
     C. Angelantonj, A. Sagnotti, {\it Type-I Vacua and Brane Transmutation},
       hep-th/0010279.}

\lref\AFIQU{G. Aldazabal, S. Franco, L. E. Ib\'a\~nez, R. Rabad\'an and 
A. M. Uranga, 
{\it D=4 Chiral String Compactifications from Intersecting Branes},
hep-th/0011073.}

\lref\KKS{S. Kachru, J. Kumar, E. Silverstein, 
{\it Vacuum Energy Cancellation in a Non-Supersymmetric String},
Phys. Rev. {\bf D59} (1999) 106004, hep-th/9807076;  \hfil\break
R. Blumenhagen, L. G\"orlich, {\it Orientifolds of 
Non-Supersymmetric Asymmetric Orbifolds}, Nucl. Phys. {\bf B551}
(1999) 601, hep-th/9812158.}

\lref\FS{W. Fischler, L. Susskind, {\it Dilaton Tadpoles,
String Condensates and Scale Invariance}, Phys. Lett. {\bf B171}
(1986) 383; {\it Dilaton Tadpoles, String Condensates and Scale
Invariance. 2}, Phys. Lett. {\bf B173} (1986) 262.} 

\lref\SM{G. Aldazabal, L. E. Ib\'a\~nez, F. Quevedo, {\it A D-brane Alternative
to the Standard Model}, hep-ph/0001083; \hfil\break
I. Antoniadis, E. Kiritsis and T.N. Tomaras,
{\it A D-brane Alternative to Unification},  hep-th/0004214; \hfil\break
G. Aldazabal, L. E. Ib\'a\~nez, F. Quevedo and A. M. Uranga, 
{\it D-Branes at Singularities : A Bottom-Up Approach to the String
Embedding of the Standard Model},JHEP 0008 (2000) 002, hep-th/0005067; 
\hfil\break
A. Krause, {\it A Small Cosmological Constant, Grand Unification and Warped
Geometry}, hep-th/0006226; \hfil\break
G. Aldazabal, S. Franco, L. E. Ib\'a\~nez, R. Rabad\'an and A. M. Uranga, 
{\it Intersecting Brane Worlds},
hep-ph/0011132.}

\lref\dumou{E. Dudas and J. Mourad, {\it Brane Solutions in
Strings with Broken Supersymmetry and Dilaton Tadpoles},
Phys. Lett. {\bf B486} (2000) 172, hep-th/0004165.} 

\lref\sugi{S. Sugimoto, {\it Anomaly Cancellations in Type I 
${\rm D9-D}\bar 9$  System and the $USp(32)$ String Theory},
Prog. Theor. Phys. {\bf 102} (1999) 685,  hep-th/9905159.}

\lref\rssop{I. Antoniadis, E. Dudas, A. Sagnotti, 
{\it Supersymmetry Breaking, Open strings and M-theory},
Nucl. Phys. {\bf B544} (1999) 469, hep-th/9807011; \hfil\break
I. Antoniadis, G. D'Appollo\-nio, E. Dudas, A. Sagnotti, {\it Partial
Breaking of Supersymmetry, Open Strings and M-theory},
Nucl. Phys. {\bf B553} (1999) 133, hep-th/9812118; 
{\it Open Descendants of $\ZZ_2 \times \ZZ_2$ Freely Acting
Orbifolds}, Nucl. Phys. {\bf B565} (2000) 123, hep-th/9907184.} 

\lref\RS{L. Randall and R. Sundrum, {\it An Alternative to Compactification},
Phys.Rev.Lett. {\bf 83} (1999) 4690, hep-th/9906064.}

\lref\typo{C. Angelantonj, {\it Non-tachyonic open descendants of the
0B string theory}, Phys. Lett. {\bf B444} (1998) 309, 
hep-th/9810214;  \hfil\break
R. Blumenhagen, A. Font, D. L\"ust, {\it Tachyon free
orientifolds of type 0B strings in various dimensions},
Nucl. Phys. {\bf B558} (1999) 159, hep-th/9904069;   \hfil\break
R. Blumenhagen, A. Kumar, {\it A note on orientifolds and
dualities of type 0B string theory}, Phys. Lett. {\bf B464} (1999) 46, 
hep-th/9906234;  \hfil\break
K. F\"orger, {\it On non-tachyonic $Z_N \times Z_M$
orientifolds of type 0B string theory}, Phys. Lett. {\bf B469} (1999)
113, hep-th/9909010;  \hfil\break
E. Dudas and  J. Mourad, {\it D-branes in non-tachyonic 0B orientifolds},
hep-th/0010179.}

\lref\kallosh{G. Gibbons, R. Kallosh and  A. Linde, {\it Brane World Sum
    Rules},  hep-th/0011225.}

\lref\flln{S. F\"orste, Z. Lalak, S. Lavignac and H.P. Nilles,
{\it The Cosmological Constant Problem from a Brane World Perspective},
JHEP {\bf 0009} (2000) 034, hep-th/0006139.} 

\lref\bhm{P. Berglund, T. H\"ubsch and D. Minic, {\it Exponential Hierarchy
from Spacetime Variable Vacua}, JHEP {\bf 0009} (2000) 015, hep-th/0005162.}     

\lref\silver{S. Kachru, M. Schulz and E. Silverstein, {\it
 Self-tuning flat domain walls in 5d gravity and string theory},
Phys.Rev. {\bf D62} (2000) 045021, hep-th/0001206.}


\Title{\vbox{
 \hbox{HU--EP--00/55}
 \hbox{hep-th/0011269}}}
{\vbox{\centerline{Dilaton Tadpoles, Warped Geometries and Large  }
       \bigskip
       \centerline{ Extra Dimensions for Non-Supersymmetric Strings } }}
\centerline{Ralph Blumenhagen\footnote{$^1$}{{\tt 
blumenha@physik.hu-berlin.de}} and Anamar\'{\i}a Font\footnote{$^2$}{{\tt 
afont@fisica.ciens.ucv.ve}. On leave of absence from Departamento de 
F\'{\i}sica, Facultad de Ciencias, Universidad Central de Venezuela. Work 
supported by a fellowship from the Alexander von Humboldt Foundation. }} 
\bigskip
\centerline{\it Humboldt-Universit\"at zu Berlin, Institut f\"ur  
Physik,}
\centerline{\it Invalidenstrasse 110, 10115 Berlin, Germany}
\smallskip
\bigskip
\centerline{\bf Abstract}
\noindent
We analyze the backreaction of dilaton tadpoles on the geometry
of non-supersymmetric strings. After finding explicit warped solutions
for a T-dual version of the
Sugimoto model, we examine the possibility of realizing
large extra dimension scenarios within the context of
non-supersymmetric string models. 
Our analysis reveals an appealing  mechanism to dynamically reduce 
the number of flat, non-compact directions in non-supersymmetric 
string theories. 

\bigskip

\Date{11/2000}

\newsec{Introduction}

Non-supersymmetric string compactifications 
\refs{\BLDI\rssop\KKS\typo\sugi\BB\URA\BGKL\AADS-\AFIQU}
have recently attracted 
attention in particular due to the possibility that in open string models
the string scale is not necessarily closely tied 
to the Planck scale \RTEV. In models containing
lower dimensional D-branes, extra large transversal directions can give rise
to a large Planck scale while leaving the string scale essentially a free 
parameter. In these large extra dimension scenarios with a string scale
in the TeV range, supersymmetry is not necessarily needed for 
protecting the gauge hierarchy. Thus, TeV strings
provide the natural arena for non-supersymmetric string models.
The last year has seen intense efforts in non-supersymmetric
string model building. Of particular interest are models with brane
supersymmetry breaking \refs{\sugi,\BB}, where the tree level bulk still 
preserves some
supersymmetry while supersymmetry is broken on the D-branes.
Phenomenological questions like the embedding of the standard model 
have been addressed in \SM, while stability issues were 
intensively discussed in \ABG.

All these models have in common that they feature a non-vanishing
dilaton tadpole, meaning that the string equations of motion are
not satisfied by  the factorized metric $M_4\times K_6$ and a constant
dilaton. In particular, for the open string models the dilaton
tadpoles appear  at disc level and their backreaction should be taken 
into account to come closer to the true quantum vacuum of the theory. 
In the deformed background the tadpole has disappeared \FS.
To find the true perturbative quantum vacuum, 
one would need to solve the string equations
of motion to all string loop levels. However, for non-supersymmetric
strings this is far beyond the reach of computational power.   

Extending the work of \dumou, we will solve the effective equations of 
motion for the string background in a prototype example.
In particular, we are interested in a simple enough toy model with D-branes
allowing transversal directions. Such a situation is provided by
a T-dual version of the Sugimoto model \sugi, where  positive
tensions are localized on two fixed points on a circle. 
Taking the backreaction of the dilaton
tadpole into account, we obtain a new class of non-trivial warped geometries 
 and dilaton profiles. In these backgrounds the expressions for    
the lower dimensional Planck scale and gauge couplings in terms of the
string scale and the internal geometry are drastically changed. 
Therefore, it is a legitimate  question whether 
these modified relations
still allow to disentangle the string scale from the Planck scale 
by increasing some radii in the internal space.

Intriguingly, the solutions we find only admit a finite lower dimensional
effective theory with zero cosmological constant, 
if the dimension of the flat, non-compact space
is not bigger than the critical value six. 
Thus, in non-supersymmetric string theories
quantum corrections can reduce the number of flat space-time dimensions,
leading to an appealing mechanism  to explain why we live in four
dimensions. 

This paper is organized as follows. In section 2 we review the construction
of the Sugimoto model and introduce the T-dual model we are going to consider
in the following.  In section 3 we present our solutions for the backreaction
of the dilaton tadpole on the geometry and the dilaton profile.
In section 4 we compute effective lower dimensional gravity and gauge
couplings and discuss the issue of large extra dimensions in these
backgrounds. In section 4 we end with some conclusions.

\newsec{Sugimoto Model}

The prototype example of an open string  model featuring 
brane supersymmetry breaking  is the so-called Sugimoto
model \sugi.
It is a non-supersymmetric version of
the Type I string. Whereas the supersymmetric Type I string
contains orientifold planes of negative tension and RR charge,
the Sugimoto model contains orientifold planes of positive 
tension and RR charge. This modification does not change the
Klein bottle amplitude at all, implying that at closed string tree
level the bulk still preserves supersymmetry.
However, in order to cancel the dangerous RR tadpole one has to introduce
32 anti-D9-branes, which of course also have positive tension. 
Thus, even though the RR charge is cancelled, the background contains
positive tension branes generating  a non-vanishing dilaton tadpole. 
Moreover, the 
M\"obius amplitude is non-vanishing, so that the model explicitly breaks
supersymmetry. Note that there does not exist any way of cancelling the
RR tadpole by a supersymmetric configuration of D-branes. 

Computing the ten-dimensional spectrum of this model, one finds
no tachyon and massless
vectors of the gauge group $USp(32)$ in addition to a massless fermion in the
antisymmetric representation. As expected from RR tadpole cancellation,
the anomalies cancel. 

The Sugimoto model already features one of the notorious problems of  
non-supersymmetric
strings, namely the presence of a dilaton tadpole, respectively a
non-vanishing cosmological constant. Thus, beyond the leading
order in the string coupling, a flat ten-dimensional
Minkowski space and a constant dilaton are not solutions of the string
equations of motion. In order to come closer to the true quantum vacuum,
in the next to leading order one should take 
the backreaction of the dilaton tadpole into account. For the 
ten-dimensional Sugimoto model with space-time filling
$\o{D9}$-branes  this has been done in \dumou, where it
was shown that the effective equations of motion admit a solution with 
less Poincar\'e  symmetry
and a non-trivial dilaton profile. More concretely, the solution found there
was a warped metric with nine dimensional Poincar\'e symmetry featuring
spontaneous compactification of the tenth direction and localization
of gravity to nine dimensions. 

In this paper we are interested in examining whether taking this backreaction
into account, it is still possible to disentangle the Planck scale from 
the string scale.  This is a non-trivial issue as the leading order
relations
\eqn\rela{  
M^2_{Pl}\sim {M_s^8\, V_d V_{6-d}\over g_s^2}, \quad\quad
{1\over g_{YM}^2}\sim {M_s^d V_d\over g_s} }
for the four dimensional Planck mass and the gauge couplings get modified.
In \rela\ $V_d$ denotes the volume longitudinal to the D-branes and
$V_{6-d}$ the volume transversal to the D-branes.

For the warped metric found in \dumou\ the four-dimensional Planck scale
and gauge coupling take the values
\eqn\dudam{  M^2_{Pl}\sim M_s^{17\over 2} \, V_5\, R_c^{3\over 2},\quad\quad
              {1\over g_4^2}\sim M_s^{11\over 2}\, V_5\, R_c^{1\over 2},}
with $R_c$ denoting the effective size of the spontaneously compactified 
direction $x_9$. With gauge coupling of order one, the relations \dudam\ imply
$M^2_{Pl}\sim M_s^3 R_c$, so that even with space-time filling
$\o{D9}$-branes a large extra dimension scenario is possible (at least
in the next-to-leading order approximation). 

In the following we will continue to investigate these quantum corrected 
space-times by studying  
a T-dual version of the Sugimoto model.
After performing one T-duality along the tenth direction, denoted $y$,
one gets a model with two positively charged O8-planes located at the two 
fixed points of the reflection $y \rightarrow -y$. 
We are cancelling the RR charge locally by
putting 16 anti-D8 branes on each fixed point, chosen to be at 
$y=L/2, 3L/2$. In one loop, $e^{0\Phi}$,   order, this appears to be a 
stable configuration.
Since the Klein-bottle and the annulus amplitude vanish, the
leading order force between two $O8$ planes, respectively two $\o{D8}$-branes
vanish. Only the M\"obius amplitude is non-vanishing and leads to an
attractive force  between an $O8$-plane and a $\o{D8}$-brane. 

Note that the resulting model
is nothing else than a non-supersymmetric version of the Type I' string. 
As in the original Sugimoto model, we are left with non-zero 
dilaton tadpoles due to the positive tension  localized at the
two fixed points. In string frame the effective action
for the metric and the dilaton is
\eqn\seffstr{
S_{S} = {M_S^8\over 2} \int d^{10}x \sqrt{-G} e^{-2\Phi}[ R + 
4 (\partial \Phi)^2]
- 32\, T \int d^{10}x \sqrt{-g} e^{-\Phi} 
[\delta(y-{\ts{L\over 2}})) + \delta(y-{\ts{3L\over 2}})] , }
where $g_{ab}=\delta_a^M \delta_b^N G_{MN}$ denotes the 9-dimensional
metric induced on the branes. We take $M,N$ to run over all
spacetime and $a,b$ over the longitudinal coordinates.
Note that the brane tension $32T$ is the same on both fixed points
and that due to the cancelled RR-Flux we can set the RR nine-form
to zero.  

\newsec{Solutions}

In this section we will construct solutions  of the equations
of motion resulting from the action \seffstr. Note that 
these equations are very similar to those encountered
in the dilatonic Randall-Sundrum scenario \silver. 
The essential difference is that we
have two branes with positive tension on a compact space. 
 
To study a more general class of solutions we first compactify 
the string theory on a $(8-D)$ dimensional torus of volume $V_{8-D}$.
Thus, in string frame we split the metric as
\eqn\mestr{
ds^2_{10,S} = ds^2_{D+2,S}+\sum_{m,n=D+3}^{10}  \delta_{mn}\, dx^m dx^n.}
We transform the resulting effective action via $G_E=e^{-{4\over D}\Phi}
G_S$ into Einstein frame to obtain
\eqn\seff{\eqalign{
S_E = &{M_S^8\, V_{8-D}\over 2} \int d^{D+2}x \sqrt{-G} \left[ R - 
{\ts{4\over D}} (\partial \Phi)^2 \right] \cr
&- 32\, T\,  V_{8-D} \int d^{D+2}x \sqrt{-g}\,  e^{{D+2\over D}\Phi}
\, \left[\delta(y-{\ts{L\over 2}}) + \delta(y-{\ts{3L\over 2}})\right] .\cr }}
The resulting equations of motion for the dilaton and the metric
are 
\eqn\geneq{\eqalign{
\partial_M(\sqrt{-G}\, G^{MN}\, \partial_N \Phi) = &{D+2\over 4}\, \lambda\, 
\sqrt{-g}\, e^{{D+2\over D}\Phi}\, [\delta(y-{\ts{L\over 2}})) + 
\delta(y-{\ts{3L\over 2}})]  \cr
&\cr
R_{MN} - \half G_{MN}\, R =& {4\over D}\left(\half G_{MN}\ G^{PQ}\, 
\partial_P \Phi
\, \partial_Q \Phi - \partial_M \Phi\, \partial_N \Phi\right)+ \cr
& \lambda g_{ab}\, \delta_M^a\, \delta_N^b\, \sqrt{{g\over G}} \, 
e^{{D+2\over D}\Phi}\,
[\delta(y-{\ts{L\over 2}})) + \delta(y-{\ts{3L\over 2}})] , }  }
where we have introduced $\lambda=32T/M_s^8$.

Due to the fact that both tensions at $y=L/2$ and $y=3L/2$ have the same sign,
there does not exist a solution to these equations
with $(D+1)$ dimensional Poincar\'e invariance.
This is in agreement with the sum rules recently derived in \refs{\flln,\kallosh}.
Therefore, the best we can try is to look for solutions with 
$D$-dimensional Poincar\'e invariance, for which we make the following
warped ansatz
\eqn\metan{
ds^2_{D+2} = e^{2M(r,y)} \eta_{\mu\nu}\, dx^\mu dx^\nu + e^{2N(r,y)} (dy)^2 
+ e^{2P(r,y)} (dr)^2 . }
We also assume that $\Phi$ depends only on $r$ and $y$. Note that
this ansatz assumes that the cosmological constant in the effective 
$D$-dimensional theory vanishes. 
As usual, one first constructs  solutions of eqs.\geneq\ in the bulk and then
imposes the jump conditions due to the branes at $y=L/2, 3L/2$.
In order to satisfy these jump conditions we are led to choose warp 
factors that depend separately on $r$ and $y$. More precisely, we take
\eqn\sepa{
M(y,r)=A(y) + X(r) \quad , \quad 
N(y,r)=B(y) + Y(r) \quad , \quad 
P(y,r)=C(y) + Z(r) . }
Likewise, we choose  $\Phi$ to depend separately on $r$ and $y$ and write
\eqn\sepp{
\Phi(y,r)=\varphi(y) + \chi(r) .}
It is straightforward to insert the above ansatz into \geneq\ to find
the equations of motion.
Simplifying the notation by introducing
\eqn\delts{\Delta =\left[\delta(y-{\ts{L\over 2}}) + 
\delta(y-{\ts{3L\over 2}})\right] ,}
the dilaton equation of motion reads 
\eqn\dil{\eqalign{
&\Bigl[\varphi^{\prime\prime} + ({\ts D}A^\prime - B^\prime + C^\prime) 
\varphi^\prime \Bigr] +\cr
& e^{2(B-C)} e^{2(Y-Z)} \Bigl[ \ddot \chi + (D\dot X - 
\dot Z + \dot Y)\dot \chi \Bigr] = 
{D+2\over 4}\, \lambda\, \exp\left(B+Y+{\ts{D+2\over D}}\,\Phi\right) \Delta, \cr}}
where primes and dots refer to derivatives with respect to $y$ and $r$
respectively.
{}From the ${\mu\nu}$ component of the Einstein equations we obtain 
\eqn\rmu{\eqalign{
&  e^{2(B-C)} e^{2(Y-Z)} \Bigl[ (D-1)\ddot X + 
{\ts{D(D-1)\over 2}}(\dot X)^2 - (D-1)\dot X\dot Z + (D-1)\dot X\dot Y
+ \cr
& \ddot Y + (\dot Y)^2 - \dot Y\dot Z + 
{\ts{2\over D}}(\dot \chi)^2 \Bigr] + \cr
\Bigl[ &(D-1)A^{\prime\prime} + {\ts{D(D-1)\over 2}}(A^\prime)^2 - 
(D-1)A^\prime B^\prime + (D-1)A^\prime C^\prime +
C^{\prime\prime} + (C^\prime)^2- \cr
&B^\prime C^\prime +
{\ts{2\over D}} (\varphi^\prime)^2 \Bigr]  
=-\lambda\, \exp\left(B+Y+{\ts{D+2\over D}}\,\Phi\right) \, \Delta.  \cr}}
The ${rr}$ component gives
\eqn\rrr{\eqalign{
&\Bigl[ DA^{\prime\prime} + {\ts{(D+1)D\over 2}}(A^\prime)^2 - 
DA^\prime B^\prime + 
{\ts{2\over D}} (\varphi^\prime)^2 \Bigr] + \cr
&  e^{2(B-C)} e^{2(Y-Z)} \Bigl[ {\ts{D(D-1)\over 2}} (\dot X)^2 + 
D\dot X\dot Y - {\ts{2\over D}}(\dot \chi)^2 \Bigr] 
= -\lambda\, \exp\left(B+Y+{\ts{D+2\over D}}\,\Phi\right)\, \Delta.  \cr}}
{} From the ${yy}$ component we find
\eqn\ryy{\eqalign{
& \Bigl[{\ts{D(D-1)\over 2}}(A^\prime)^2 + DA^\prime C^\prime - 
{\ts{2\over D}} (\varphi^\prime)^2 \Bigr]  +\cr
& e^{2(B-C)} e^{2(Y-Z)} \Bigl[ D\ddot X + {\ts{(D+1)D\over 2}}
 (\dot X)^2 - D\dot X\dot Z + 
{\ts{2\over D}}(\dot \chi)^2 \Bigr] =0, \cr}}
and finally the off-diagonal ${yr}$ equation has the form
\eqn\ryr{\eqalign{
&D(A^\prime \dot X - A^\prime\dot Y - C^\prime \dot X) +
{\ts{4\over D}}\, \varphi^\prime \dot \chi=0.    }}
Notice that the jump conditions implied by eqs. \dil, \rmu\ and \rrr\
require discontinuous $\varphi^\prime$, $A^\prime$ and $C^\prime$.
Apparently, consistent solutions must satisfy
\eqn\scon{\dot Y + {\ts{D+2\over D}} \dot \chi =0 , }
so that there is no $r$ dependence on the right hand side of the
jump conditions. Substituting \scon\ into  \ryr\ gives
\eqn\nscon{
D^2(A^\prime - C^\prime) \dot X + 
((D+2)D A^\prime + 4\varphi^\prime) \dot \chi=0    ,}
which severely restricts the solutions.

By rescaling the coordinates $r$ and $y$ we are free to choose
$B=C$, $Y=Z$. It turns out that
most of the possible solutions of \nscon\ are inconsistent with the
remaining equations. For instance, taking $\dot X=0$ which requires
$\dot \chi =0$ or $4\varphi^\prime = - D(D+2) A^\prime$, leads 
to bulk solutions incompatible with the brane matching conditions. 
Similarly, the choice $A=C$ and $4\varphi^\prime = - D(D+2)A^\prime$ 
is inconsistent. Altogether, we find two consistent solutions, 
labelled I and II and discussed in the following.

\noindent
\subsec{Solution I}
\noindent 
We take $A=C$ and satisfy \nscon\ by fixing $\dot \chi=0$. Notice that 
we can set $Y=0$. The eqs.  \rmu, \rrr\ and \ryy\
 imply $\dot X = \pm K$, for some positive
constant $K$. For concreteness we choose $\dot X = - K$.
The remaining equations simplify to
\eqn\soli{\eqalign{
\varphi^{\prime\prime} + DA^\prime  \varphi^\prime
&= {\ts{D+2\over 4}}\lambda\, 
\exp\left(A+{\ts{D+2\over D}}\,\Phi\right)\, \Delta \cr
A^{\prime\prime} + D(A^\prime)^2  + D\, K^2 
&= -{\ts{\lambda\over D}}\, \exp\left(A+{\ts{D+2\over D}}\,\Phi\right)\,
 \Delta \cr  
(A^\prime)^2 - {\ts{4\over D^2(D+1)} }(\varphi^\prime)^2 
+ K^2 &=0 .} }
In the bulk these equations are solved by
\eqn\bulki{\eqalign{
A(y) &= {\ts{1\over D}} \log \Bigl|\,\sin\bigl[DKy + 
2\theta\bigr]\Bigr| \cr
\Phi_{\pm}(y) &= \pm {\ts{\sqrt{D+1}\over 2}}\, \log\Bigl|\,\tan
\bigl[{\ts{D\over 2}}Ky + \theta\bigr]\Bigr| + \Phi_0 \cr
X(r)&=-Kr, }}
where in $A$ we have dropped a constant that can be absorbed in rescaling
the coordinates $x_\mu$.
\noindent
In order to solve the jump conditions we choose
\eqn\ji{\eqalign{
A(y) &= {\ts{1\over D}} \log \Bigl|\,\sin\bigl[DKy \bigr]\Bigr| 
 \quad , \quad 0 \leq y \leq {\ts{L\over 2}} \cr 
A(y) &= {\ts{1\over D}} \log \Bigl|\,\sin\bigl[DK(y-L) \bigr]\Bigr| 
\quad , \quad {\ts{L\over 2}} \leq y \leq L ,}}
and similarly for $\varphi$. In the interval $[L,2L]$ the solution is
extended periodically. Matching then requires
\eqn\cfixi{ \cos \left[{\ts{D\over 2}}KL\right] = 
\mp {{2\sqrt{D+1}\over D+2}}
\quad\quad ; \quad\quad
e^{{D+2\over D}\Phi_0} = {K\over {\eta(D) \lambda}} ,}
where $\eta(D)$ is a numerical factor that can be easily evaluated,
{\it e.g.} $\eta(8)=0.30$ and $\eta(4)=0.51$.
We choose ${D\over 2}KL<\pi/2$, 
and correspondingly $\Phi_-$, so that the
metric has only singularities at $y=0,L$. 

By computing the Ricci scalar
in string frame one finds divergences at $y=0,L$, so that,
similar to \bhm, we have naked
singularities in the internal space. However, the dilaton also
diverges at these points leading to an infinite string coupling at the 
singularity. Thus, our next-to-leading order treatment of the
string loop expansion breaks down and one might hope that higher
loop or non-perturbative effects cure this singularity. 
After all it is not too surprising that we find these singularities
in the solution. Roughly speaking, developing these singularities is the way
gravity can handle a configuration of sources (two positive tensions) 
that for RR-fields (two positive RR-charges) would be inconsistent.

In the resulting metric with $D$-dimensional Poincar\'e
invariance the coordinate $r$ is also non-compact.
In general, we can define an effective size for this coordinate
\eqn\rsize{
\rho = \int_{-\infty}^\infty dr \, e^{Z_s} , }
where $Z_s=Z+{2\over D}\chi$ is the corresponding warp factor in
the string metric. Since in this solution $Z=0$ and $\dot \chi=0$, 
we see that $\rho$ is unbounded.

Note that in contrast to our situation, in the Randall-Sundrum (RS) set up \RS\
an exponential warp factor led to localized gravity in one dimension
lower. In the RS case this was due to the choice $X(r)=-K|r|$, which 
is not allowed in our case, as it introduces a new singularity 
in the $r$ direction. 

\noindent
\subsec{Solution  II}

\noindent
This solution will turn out to be much more  interesting and non-trivial
than solution I.  
We only assume $\dot X = \a \dot \chi$, with $\a$  
constant as required by eq. \nscon. By virtue of variable
separation, the bulk equations reduce to
\eqn\solii{\eqalign{
A^{\prime\prime} + D\, (A^\prime)^2  &= -D\, K^2 \cr
\varphi^{\prime\prime} + D\, A^\prime \varphi^\prime  &= -D\, K^2/\a \cr
C^{\prime\prime} + D\, A^\prime C^\prime &= -D\, \mu K^2  \cr
\ddot \chi + D\, \dot X \dot \chi &= D\, K^2/\a, }}
where $K$ and $\mu$ are constants. There are further relations
\eqn\auxii{\eqalign{
2D\, A^\prime C^\prime - (D-1)\, A^{\prime\prime} - {\ts{4\over D}} 
(\varphi^\prime)^2
&=-D\, K^2\left( \mu + 1 + {\ts{D+2\over \a\,D}} \right)\cr
2D\, \dot X \dot Y - (D-1) \ddot X - {\ts{4\over D}} (\dot \chi)^2
&=-D\, K^2 \left( \mu - 1 + {\ts{D+2\over \a\,D}} \right) .} }
After some more computational steps, the solutions to the bulk equations 
turn out to be 
\eqn\bii{\eqalign{
A(y) &= {\ts{1\over D}} \log \Bigl|\,\sin\bigl[DKy + 
2\theta\bigr]\Bigr| \cr
C_{\pm}(y) &= {\ts{\mu\over D}} \log \Bigl|\,\sin\bigl[DKy + 
2\theta\bigr]\Bigr| \pm
{\ts\sqrt{8\over D}\, {1\over {\a D}}} \log\Bigl|\,\tan
\bigl[{\ts{D\over 2}}Ky + \theta\bigr]\Bigr| \cr
\Phi_{\pm}(y,r) &={\ts{1\over {\a D }}} \log \Bigl|\,\sin\bigl[DKy + 
2\theta\bigr]\Bigr| \pm
{\ts\sqrt{ D\over 2}} \log \Bigl|\,\tan\bigl[{\ts{D\over 2}}Ky+ 
\theta\bigr]\Bigr| + \cr
&\ \  {\ts{1\over {\a D}}} \log\Bigl|\,\cosh \bigl[DKr + 
\beta\bigr]\Bigr| + \Phi_0\cr
X(r) &={\ts {1\over D}} \log \Bigl|\,\cosh \bigl[DKr + 
\beta\bigr]\Bigr| \quad ; \quad Y(r)= -{\ts{D+2\over {\a D}}} X(r).
}}
Substituting into eqs. \nscon\ and \auxii\ determines the constants
$\a$ and $\mu$ to be
\eqn\const{ \mu = {D+1 \over 2} + {2\over {\a^2 D^2 }}, \quad  \quad
\a _{\pm}= {(D+2) \pm \sqrt{(D+8)D}\over D(D-1)} .}
We have absorbed an integration constant in $C_\pm$ in a redefinition
of $K$. 
As we will explain, the solution with $\a_-$ 
leads to diverging lower dimensional quantities, so that in the following 
we discuss only the case $\a_+$.

To solve the matching relations we again choose $A$ of the form
\ji\ and similarly for $C$ and $\varphi$. Remarkably, the three jump 
conditions  turn out to be compatible and lead to 
\eqn\cfixii{\cos \left[{\ts{D\over 2}}KL\right] = \mp \sqrt{8\over D+8}, 
\quad\quad e^{{D+2\over D}\Phi_0} = {K\over \kappa(D)\lambda} ,}
where the sign corresponds to the free sign in
$C_{\pm}$ and $\Phi_{\pm}$. We choose ${D\over 2}KL<\pi/2$ so that the 
metric, as well as the dilaton, have singularities only at $y=0,L$.
Thus, we choose $\Phi_-$ and $C_-$ in \bii.
The numerical coefficients $\kappa(D)$ can be found in Table 1.
Consistently, the right hand side of the first equation in \cfixii\
is smaller than one.
The qualitative form of the solutions for $A,C$ and $\varphi$ are shown 
in figures 1-3.

\noindent
\vbox{
\hbox{\noindent\epsfysize=7truecm\epsfbox{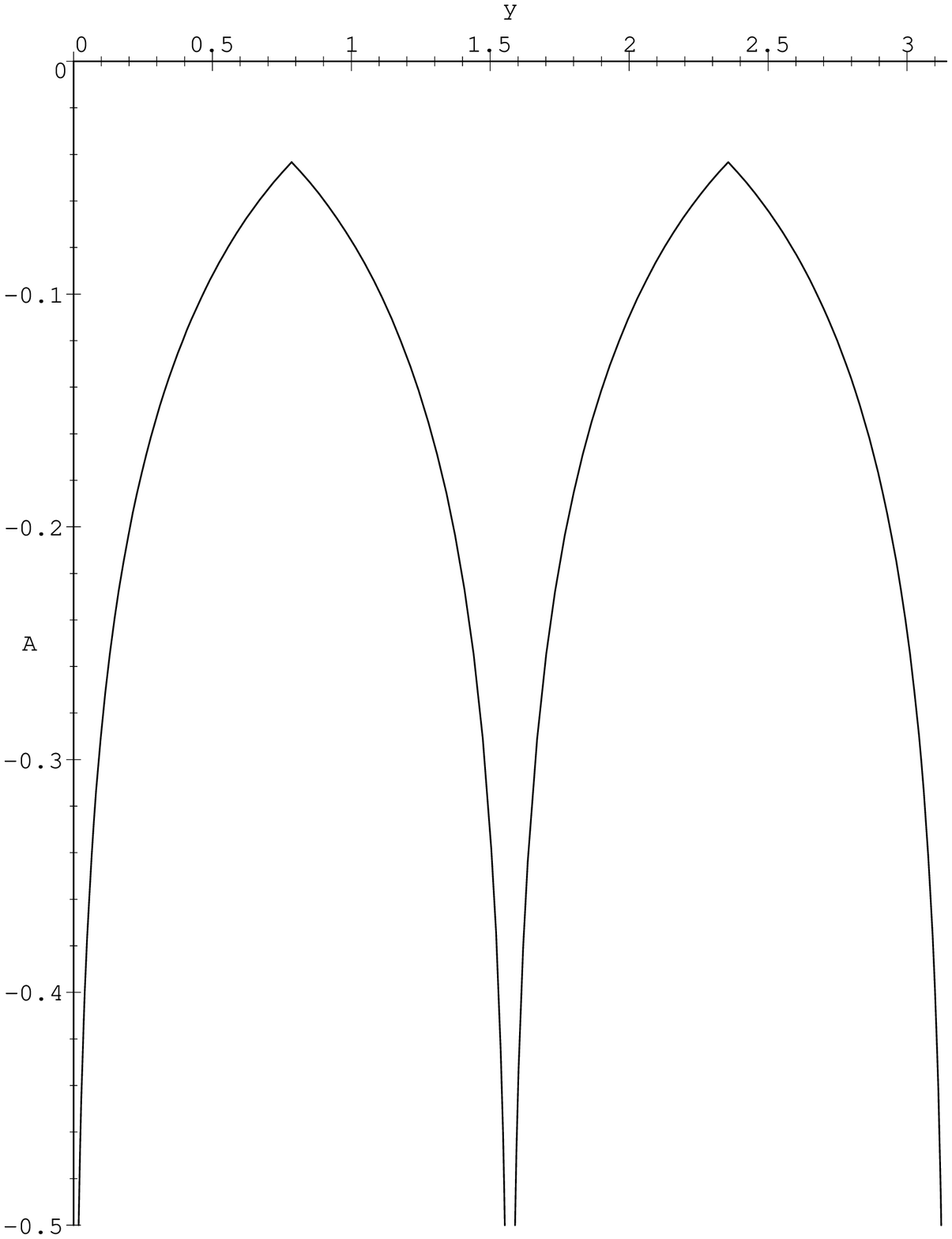}\hskip 0.5cm
\epsfysize=7truecm\epsfbox{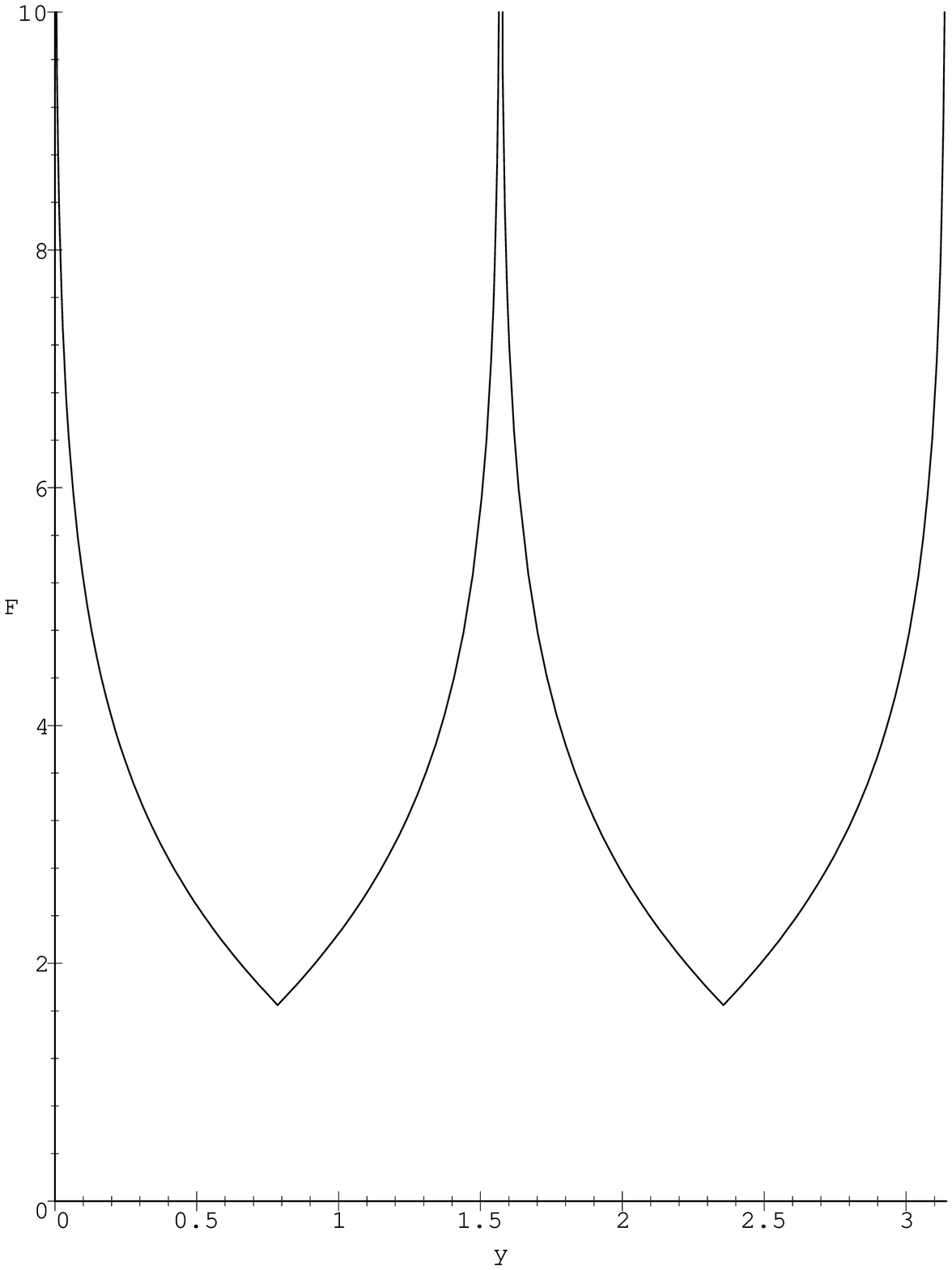} \hskip 0.5cm
\epsfysize=7truecm\epsfbox{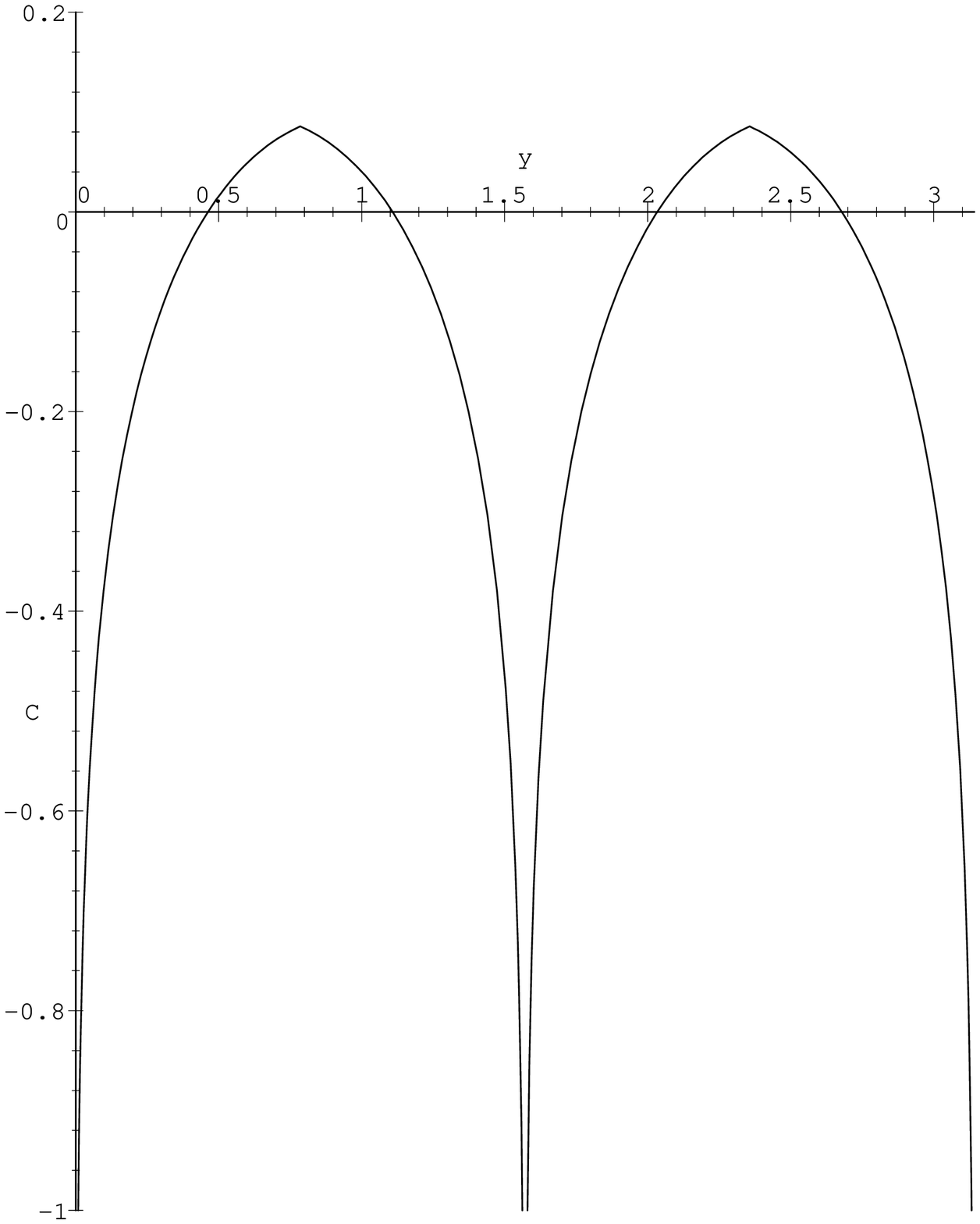}}
\noindent
\hbox{\hskip 1.2cm Fig.1: $A(y)$ \hskip 3.2cm Fig.2: $\varphi(y)$ 
      \hskip 3.2cm Fig.3: $C(y)$}}
\bigskip\noindent
In this case the coordinate $r$ is actually compact.
Computing the effective size according to \rsize\ gives
\eqn\rsizev{
\rho = \int_{-\infty}^\infty dr\, e^{Z_{S}} = \epsilon(D)\, L, }
where the numerical coefficients can be found in Table 1.
In contrast to the tree level result, the sizes of the
two non-flat directions are correlated.  

As in solution I there appear naked singularities and diverging string
couplings at $y\in\{0,L\}$. The comments made for solution I also apply
here. Moreover, we find  singularities for $r\to\pm\infty$, 
where also the string coupling diverges.

\newsec{Effective couplings}

Even though at string tree level we compactified only the direction $y$, the
backreaction of the dilaton tadpoles forced us to spontaneously
compactify another direction $r$. Thus, we do not get an effective
theory with $(D+1)$ dimensional Poincar\'e invariance. The best 
we can hope for is an effective theory with $D$ dimensional
Poincar\'e symmetry. This of course is very similar to the Randall-Sundrum 
scenario, where gravity confines to some lower dimension simply
by a non-trivial warp factor.
Thus, given the solutions found in the previous section we now want to analyze
whether gravity and gauge interactions are really confined to the
$D$-dimensional space-time. To this end we compute the $D$-dimensional
Planck mass and gauge couplings. After transforming to the Einstein
frame these quantities are given by
\eqn\mplg{\eqalign{
M_{Pl}^{D-2} & = M_s^8\, V_{8-D} \int_0^{2L} dy \int_{-\infty}^\infty 
dr \, e^{(D-2)A + B + C + (D-2)X + Y + Z}  \cr
{1\over{g_{D}^2}} & = M_s^5\, V_{8-D} \int_0^{2L} dy 
\int_{-\infty}^\infty dr \,
e^{{(D-6)\over D}\Phi} \, e^{(D-4)A + C  + (D-4)X  + Z}
[\delta(y-{\ts{L\over 2}})) + \delta(y-{\ts{3L\over 2}})] . } }
In the case of solution I, for both quantities the integral in $r$ 
diverges as  the only $r$ dependence appears in  $X=-Kr$.
Therefore, the solution I does not lead to a finite effective theory in
$D$-space-time dimensions.

Contrarily, in the case of solution II, $M_{Pl}$ turns out to be 
finite provided we choose $\a_+$. More concretely, by evaluating the integrals
numerically we obtain
\eqn\mpii{
M_{Pl}^{D-2}  = \gamma(D)\,{M_s^8 V_{8-D}\over {K^2}},}
where the numerical coefficients are given in Table 1.
On the other hand, for the  Yang-Mills coupling we obtain
\eqn\ymcou{ {1\over {g_{D}^2}}=\delta(D)\,
{M_s^5\, V_{8-D} \over K}\,  e^{{(D-6)\over D}\Phi_0}.}
As can be seen from Table 1, the coefficient $\delta(D)$ diverges for
$D\in\{8,9\}$ and is finite only for $D\le 6$.

\vskip 0.8cm
\centerline{\vbox{
\hbox{\vbox{\offinterlineskip
\def\tablespace{height2pt&\omit&&\omit&&\omit&&\omit&&
 \omit&\cr}
\def\tablerule{\tablespace\noalign{\hrule}\tablespace}

\hrule\halign{&\vrule#&\strut\hskip0.2cm\hfil#\hfill\hskip0.2cm\cr
\tablespace
& $D$ && $\kappa(D)$ && $\epsilon(D)$ && $\gamma(D)$ && $\delta(D)$ &\cr
\tablerule
& 2    && 1.41 && 19.73 && 0.74  &&  0.14  &\cr
\tablespace
& 3    && 1.04 && 10.98 && 0.72  &&  0.33  &\cr
\tablespace
& 4    && 0.85 && 8.07  && 0.70  &&  0.67  &\cr
\tablespace
& 5    && 0.73 && 6.60  && 0.68  &&  1.44  &\cr
\tablespace
& 6    && 0.64 && 5.72  && 0.67  &&  5.87  &\cr
\tablespace
& 7    && 0.58 && 5.12  && 0.66  &&  $\infty$  &\cr
\tablespace
& 8   && 0.53 && 4.70  && 0.65  &&  $\infty$  &\cr
\tablespace}\hrule}}}}
\centerline{
\hbox{{\bf Table 1:}{\it ~~ Numerical coefficients.}}}
\vskip 0.5cm 

Thus, we only get a bona fide effective theory with at most six-dimensional 
Poincar\'e symmetry. This is in contrast to supersymmetric vacua, where
the number of flat directions is a free parameter. 
We conclude, that in non-supersymmetric theories
the number of flat non-compact directions is not a free parameter, but
can be restricted by the dynamics. This hints to an appealing dynamical 
mechanism to  explain why we live in four dimensions. 

Finally, let us see whether the solution admits to disentangle the
Planck and the string scale. 
After a further toroidal compactification on $T^{(D-4)}$ 
to four flat dimensions,  we obtain  the 
following relations for the four dimensional scales
\eqn\lowrel{  M^2_{Pl} \sim M_s^8 \, V_{8-D}\, W_{D-4}\, L^2,
\quad\quad
              {1\over g_4^2} \sim\, M_s^{(4D+16)\over D+2}\, 
        V_{8-D}\, W_{D-4} \, L^{8\over D+2}, }
where $W_{D-4}$ is the volume of $T^{(D-4)}$. 
Note that these relations differ from the tree level results \rela.
Choosing the gauge coupling of order one  implies
\eqn\disn{     M^2_{Pl} \sim M_s^{4D\over D+2}\, L^{(2D-4)\over D+2},}
showing that $M_s$ is a free parameter as long as we choose
the radius $L$ large enough. We conclude, that large extra dimension scenarios
are possible even when the next to leading order quantum corrections
to the background are taken into account. 
Inserting numerical values into \disn\ and choosing $M_s={\rm 1TeV}$ gives
the rough estimates of the internal dimensions shown in Table 2.  
\vskip 0.8cm
\vbox{\centerline{\vbox{
\hbox{\vbox{\offinterlineskip
\def\tablespace{height2pt&\omit&&\omit&&
 \omit&\cr}
\def\tablerule{\tablespace\noalign{\hrule}\tablespace}

\hrule\halign{&\vrule#&\strut\hskip0.2cm\hfil#\hfill\hskip0.2cm\cr
\tablespace
& $D$ && $L$ && $(V_{8-D}\, W_{D-4})^{1\over 4}$ &\cr
\tablerule
& 6    && $10^{14}$m &&  $10^{-27}$m  &\cr
\tablespace
& 5    &&  $10^{19}$m && $10^{-30}$m  &\cr
\tablespace
& 4    && $10^{30}$m &&  $10^{-35}$m    &\cr
\tablespace}\hrule}}}}
\centerline{
\hbox{{\bf Table 2:}{\it ~~ Large extra dimensions.}}}}
\vskip 0.5cm 
Thus, in agreement with  the naive tree-level result \rela,
in order to obtain phenomenologically acceptable sizes one has to
apply more T-dualities to get D-branes with more transversal directions.
However, extrapolating the results presented in \dumou\ and in the present
paper, it is a non-trivial question whether the critical dimension
for such solutions would be larger than three. 

\newsec{Conclusions}

In this paper we have studied the backreaction of the dilaton tadpole 
for a prototype model featuring brane
supersymmetry breaking. Making a warped ansatz for the metric we have found
a highly non-trivial solution to the equations of motion, which allowed
to compute finite effective couplings of a lower dimensional
theory in flat space provided the dimensions were smaller than the critical
value  six.  We have shown that the naive tree level relations
for these couplings in terms of the string scale and the internal
geometry change, but that they in principle still allow large extra dimensions
and a string scale in the TeV range. However, for $\o{D8}$ branes
these extra dimensions were unacceptable large, so that one should
study models with lower dimensional branes.
It would also be interesting to determine the spectrum in these
highly curved backgrounds and in particular discuss stability issues.

\vskip 1cm

\centerline{{\bf Acknowledgements}}\pano
This work  is supported in part by the European Comission
RTN programme HPRN-CT-2000-00131 and the Alexander von Humboldt Foundation.
R.B. also thanks the Aspen Center for Physics, where part of this 
work was performed. 
Moreover, we would like to thank K. Behrndt, E. Dudas and 
R. Tatar for  discussion, and D. L\"ust for useful comments and 
reading the manuscript

\vskip 1cm

\listrefs

\bye